# Estimating the Aggregate/Intraaggregate Mass Ratio of a Shrinking Soil

V.Y. Chertkov*

Agricultural Engineering Division, Faculty of Civil and Environmental Engineering, Technion, Haifa 32000, Israel

**Abstract:** A recently introduced parameter, the ratio of an aggregate solid mass to a solid mass of an intraaggregate matrix ($K$ ratio) is connected with the mean thickness of a deformable, but non-shrinking surface layer of aggregates and is a fundamental property of aggregated soils that essentially influences their shrinkage. The objective of this work is to suggest and validate an approach to estimating the $K$ ratio at any soil clay content through characteristics of soil texture and structure. We derive an equation that reflects the interrelation between the $K$ ratio and soil texture and structure. The $K$ ratio can be estimated as the solution of the equation and is determined by the mean size of soil solids and the maximum size of soil aggregates in the oven-dried state, independently of a measured shrinkage curve. To validate the approach we use available data for eight soils.

## INTRODUCTION

A recently proposed model [1, 2] showed how a clay shrinkage curve under the influence of a silt-sand admixture as well as inter- and intraaggregate structures, is transformed into the soil shrinkage curve at any soil clay content. In general, the shrinkage curve of a soil is non-single valued since the crack volume contribution to the shrinking soil volume depends on sampling, sample preparation, sample size, and drying regime. The above model, unlike previous ones (*e.g.*, [3-6]) investigated the reference shrinkage curve that, by definition, corresponds to shrinkage without interaggregate cracking and can be predicted in a single valued manner. In addition, the previous models justify some *a priori* taken mathematical expression for a shrinkage curve (different for each different approach) by the fitting of its parameters (from 3 to 11 depending on the approach) to experimental shrinkage curve data. Unlike these models, Chertkov [1, 2] derived the reference shrinkage curve of a soil from the shrinkage curve of a clay [7, 8] contributing to the soil and two new features of the intraaggregate soil structure (Fig.**1**): (i) the existence and dewatering of a deformable, but non-shrinking aggregate surface layer (interface layer); and (ii) the existence and volume increase of intraaggregate lacunar pores at soil shrinkage. The interface layer exists at any soil clay content (Fig.**1**). Its mean thickness increases with a clay content decrease. The lacunar pores exist at a clay content, $c$ lower than a critical value, $c^*$ (Fig.**1**).

The derivation of the reference shrinkage curve explains the transition from a clay shrinkage curve to a soil shrinkage curve and leads to the understanding of the origin of the shape of a soil shrinkage curve. In addition, the reference shrinkage curve can be used in estimating the contribution of the crack volume to the soil shrinkage curve, the soil hydraulic properties, and in other applications.

The methodology of the reference shrinkage curve prediction was based on detailed accounting for possible contributions to the soil volume and water content during shrinkage. The reference shrinkage curve is determined by eight physical soil parameters: oven-dried specific volume ($Y_z$), maximum swelling water content ($W_h$), mean solid density ($\rho_s$), soil clay content ($c$), oven-dried structural porosity ($P_z$), the ratio of aggregate solid mass to solid mass of intraaggregate matrix ($K$), the lacunar factor that characterizes the rate of the lacunar pore volume change with water content ($k$), and oven-dried lacunar pore volume ($U_{lpz}$). The model [1, 2] was validated using data for six soils from [9], a soil from [10], and eight soils from [11].

The last three parameters ($K$, $k$, and $U_{lpz}$) present new fundamental physical soil properties and as such, in principle, can be measured or estimated from some measurements independently of a measured shrinkage curve. The methods of estimating $k$ and $U_{lpz}$ were briefly indicated [2]. Estimating the $K$ ratio independently of a measured shrinkage curve in the particular case of the thin interface layer (Fig.**1** at $c>c^*$ when the lacunar pores are lacking) was noted without validation [12]. The objective of this work is to propose and validate an approach for estimating the $K$ ratio in the general case of any soil clay content, from the oven-dried aggregate-size distribution and soil texture. Notation is summarized at the end of the paper.

---

*Address correspondence to this author at the Agricultural Engineering Division, Faculty of Civil and Environmental Engineering, Technion, Haifa 32000, Israel; E-mail: agvictor@tx.technion.ac.il

# THEORY

**General Expression for the *K* Ratio**

The *K* ratio can be written as [1, 2]

$$K=(1-U_i/U_h)^{-1} \qquad (1)$$

where $U_i$ is the specific volume of the deformable, but non-shrinking aggregate surface layer (interface layer) (Fig.**1**), and $U_h$ is the maximum specific volume of the intraaggregate matrix (Fig.**1**) when shrinkage starts (at the total water content $W=W_h$). Each aggregate has the interface layer (Fig.1) that contributes to $U_i$. This means that $U_i$ can be expressed through the aggregate-size distribution at the total water content $W=W_h$ when shrinkage starts as

$$U_i = \frac{(1-P_h)}{\rho_b} \int_{X_{min}}^{X_m} \frac{V_i(X,x)}{V_a(X)} \frac{dF(X,P_h)}{dX} dX . \qquad (2)$$

In Eq.(2) $F(X,P_h)$ is the cumulative aggregate-size distribution as a function of the current aggregate size, $X$ and structural porosity, $P_h$ of the soil at $W=W_h$; $V_i(X, x)$ is the volume of the interface layer of the mean thickness, $x$ relating to a separate aggregate of size $X$ (Fig.**1**); $V_a(X)$ is the total volume of a separate aggregate of size $X$; $X_{min}$ and $X_m$ are the minimum and maximum sizes of aggregates, respectively, at $W=W_h$; $\rho_b$ is the dry bulk density of the soil (the soil solid mass per unit soil volume at $W=W_h$).

$(dF(X,P_h)/dX) \cdot dX$ in Eq.(2) gives a volume of aggregates in the size range $dX$ per unit volume of aggregates of all the possible sizes, $X_{min} \leq X \leq X_m$ (at $W=W_h$). $(1/V_a(X))(dF(X,P_h)/dX) \cdot dX$ in Eq.(2) gives the number of aggregates in the size range $dX$ per unit volume of all aggregates (at $W=W_h$). $(V_i(X, x)/V_a(X))(dF(X,P_h)/dX) \cdot dX$ in Eq.(2) gives the interface layer volume (of the mean thickness $x$) relating to aggregates in the size range $dX$ per unit volume of all aggregates (at $W=W_h$). Integration over $X$ gives the total interface layer volume (of the mean thickness $x$) of all aggregates per unit volume of the aggregates (at $W=W_h$). Multiplication by $(1-P_h)$ gives the total interface layer volume of all aggregates per unit volume of the soil including structural pores (at $W=W_h$). Finally, division by $\rho_b$ gives the total interface layer volume per unit mass of oven-dried soil, that is, the specific volume $U_i$ of the interface layer.

Replacement of $U_i$ in Eq.(1) with $U_i$ from Eq.(2) leads to the general expression for the *K* ratio. In the following sections we specify the different values entering the *K* ratio expression from Eq.(2).

**Transformation of the Bulk Density Term**

One can see that $U_h/(1-P_h)$ is the maximum specific soil volume at $W=W_h$ when the interface layer and intraaggregate matrix (Fig.**1**) have the same water content and specific volume [1, 2]. This means that the multiplier $(1-P_h)/(\rho_b U_h)$ before the integral in the above expression for the *K* ratio is reduced to unity

$$(1-P_h)/(\rho_b U_h)=1 . \qquad (3)$$

**Approximation of Aggregate Shape**

The aggregate-size distribution, $F(X,P_h)$ (see the following section), as any size distribution (of pores, grains, aggregates, and others), by definition, only accounts for the size of aggregates ($X$), but not their shape. Therefore, considering the $V_i(X,x)/V_a(X)$ ratio from Eq.(2) we can in the same approximation neglect the effect of the aggregate shape on its volume ($V_a$) and volume ($V_i$) of its interface layer of the thickness $x$ (Fig.**1**). This means that when specifying the $V_i/V_a$ ratio in Eq.(2), we can consider in the same approximation the aggregates of the simplest shape, that is, the spheres of diameter $X$. One can easily be convinced that for a sphere of diameter $X$ and its surface layer of thickness $x$ the $V_i/V_a$ ratio should be

$$V_i(X, x)/V_a(X)=1-(1-2x/X)^3 . \qquad (4)$$

In addition, the same $V_i/V_a$ ratio holds for a cube of side $X$ and its surface layer of thickness $x$. In general, the (mean) interface layer thickness, $x$ can have any value in the range $0<x<X/2$ depending on the soil clay content. In the case of the clay content higher than critical ($c>c^*$) we deal with the relatively thin interface layer ($x<<X/2$) when the *K* ratio is in the range $1<K \leq \cong 1.3$ [1]. In this case Eq.(4) gives $V_i/V_a \cong 6x/X$. Estimation of the *K* ratio at $c>c^*$ (without experimental validation) was briefly considered [12]. At a clay content lower than critical ($c<c^*$) the non-shrinking interface layer can even nearly occupy the total volume of an aggregate ($x \to X/2$) when the clay content is sufficiently small, in this case $K \to \infty$ [2].

**Aggregate-Size Distribution**

Based on the intersecting surfaces approach to soil structure [13] we use in Eq.(2) the simplest aggregate-size distribution, $F(X, P_h)$ as

$$F(X, P_h) \equiv F(\eta, P_h) = (1-P_h^{I_o(\eta)/8.4})/(1-P_h) \qquad (5)$$

where

$$\eta \equiv (X-X_{min})/(X_m-X_{min}), \qquad X_{min} \leq X \leq X_m \qquad (6)$$

and

$$I_o(\eta) = \ln(6)(4\eta)^4 \exp(-4\eta) . \qquad (7)$$

In the particular, but important, case of an aggregated soil with negligible structural porosity ($P_h \to 0$) the $F(X, P_h)$ distribution in Eq.(5) should be replaced with [13]

$$F(\eta, 0) = \{1-\exp[-I_o(\eta)]\}/[1-\exp(-8.4)] . \qquad (8)$$

**Structural Porosity**

The initial structural porosity, $P_h$ (*i.e.*, at $W=W_h$) that enters the $F(X,P_h)$ distribution (Eq.(4)) can be calculated as [12]

$$P_h = [1+\{[(u_h/u_s)/(\rho_s K)+U_i]/[(u_z/u_s)/(\rho_s K)+U_i]\}(1-P_z)/P_z]^{-1}, \qquad 0 \leq W \leq W_h \qquad (9)$$

where $u_s$ is the relative volume of solids in the intraaggregate matrix; $u_z$ is the relative oven-dried volume of the intraaggregate matrix; and $u_h$ is the relative volume of the intraaggregate matrix at $W=W_h$. The values of $u_s$, $u_z$, and $u_h$ as well as $U_i$ and, therefore, $P_h$ (Eq.(9)) are calculated through the above input parameters: $Y_z$, $W_h$, $\rho_s$, $c$, $P_z$, $K$, $k$, and $U_{lpz}$ [1, 2]. Thus, $P_h$ is a function of the $K$ ratio (at given parameters $Y_z$, $W_h$, $\rho_s$, $c$, $P_z$, $k$, and $U_{lpz}$).

**The Relation between the Mean Thickness of an Interface Layer (*x*) and Soil Texture**

The soil of the interface layer and that of an intraaggregate matrix (Fig.**1**) are characterized by the same texture, *i.e.*, by the same weight fractions of clay particles, silt, and sand grains. We believe that the mean thickness of the interface layer, at least by order of magnitude, is directly connected with a certain characteristic size of internal non-homogeneities of the intraaggregate soil matrix (Fig.**1**). The natural (and single) pretenders for the role of such non-homogeneities (at the scale we consider, Fig.**1**) are different solids (clay particles, silt, and sand grains), and the single natural candidate for the role of such characteristic size is the mean size of the soil solids (non-homogeneities) by their weight fractions, $x_n$. Thus, we accept the natural assumption - that the doubled mean interface layer thickness ($2x$) is approximately equal to the mean size of the soil solids ($x_n$):

$$2x \cong x_n . \qquad (10)$$

At high clay content ($c \to 1$) $x_n$ strives to a clay particle size, and the interface layer (of the $x$ thickness; Fig.**1**) is very thin. That is, $K \to 1$. At small clay content ($c << c^*$) the mean size of soil solids, $x_n$ strives to a silt-sand grain size (depending on the grain-size distribution). With that aggregates are also degenerated to rigid silt and sand grains. This means that the intraaggregate matrix (Fig.**1**) disappears, and the aggregate /intraaggregate mass ratio, $K \to \infty$.

**Minimum and Maximum Sizes of Aggregates**

$X_{min}$ and $X_m$ in Eq.(2) and (6) relate to $W=W_h$ and can be expressed through $X_{minz}$ and $X_{mz}$ values in oven-dried state, respectively. At a given water content the intraaggregate matrix volume is proportional both to the cube of the difference between the current aggregate size and doubled interface layer thickness, $(X(W)-2x)^3$ (Fig.**1**) and to the relative volume of the intraaggregate matrix, $u(W)$. In particular, for the maximum aggregate size, $X_m$ at $W=W_h$ and that, $X_{mz}$ at $W=0$ one can write

$$(X_m - 2x)^3 \propto u(W_h) \equiv u_h \tag{11}$$

and

$$(X_{mz} - 2x)^3 \propto u(0) \equiv u_z \;, \tag{12}$$

respectively. It follows that

$$X_m = 2x + (X_{mz} - 2x)(u_h/u_z)^{1/3} \;. \tag{13}$$

The above consideration is applicable to any aggregate size. Replacing in Eq.(11)-(13) the maximum aggregate size values, $X_m$ and $X_{mz}$ with the minimum aggregate size values, $X_{min}$ and $X_{minz}$, respectively, we obtain

$$X_{min} = 2x + (X_{minz} - 2x)(u_h/u_z)^{1/3} \;. \tag{14}$$

The maximum aggregate size in the oven-dried state ($X_{mz}$) can essentially exceed the doubled mean interface layer thickness ($2x$) (Fig.**1**). Unlike that the difference between $X_{minz}$ and $2x$ in Eq.(14) in any case should be relatively small. That is $(X_{minz}-2x)/(2x) \ll 1$. Indeed, any current aggregate size ($X$) of a shrinking soil should exceed the doubled mean interface layer thickness ($2x$) (Fig.**1**; $X > 2x$). Otherwise, aggregates would turn into rigid formations. That is, they would be degenerated to silt and sand grains. It follows that the minimum aggregate size [at any water content; *e.g.*, both $X_{min}$ and $X_{minz}$, see Eq.(14)] should be close to $2x$ (but exceeding that). Thus, Eq.(14) can be rewritten as

$$X_{min} \cong X_{minz} \cong 2x \;. \tag{15}$$

Equations (10) and (15) lead to a simple, but important structural relation between the minimum aggregate size ($X_{min}$) and the mean size of the soil solids - non-homogeneities ($x_n$) as

$$X_{min} \cong X_{minz} \cong x_n \tag{16}$$

In addition, Eq.(10) and (13) present the maximum aggregate size ($X_m$) as

$$X_m = x_n + (X_{mz} - x_n)(u_h/u_z)^{1/3} \tag{17}$$

It is worth noting that Eq.(16)-(17) relate to any soil, and permit one to trace the transition from a shrinking to a rigid soil. Indeed, if the shrinkage of a soil is negligible (the clay content, $c \to 0$) the ratio, $u_h/u_z \to 1$ and Eq.(17) gives $X_m \to X_{mz}$ as it should be for a rigid soil. In addition, in this case ($c \to 0$) $x_n$ is reduced to the mean size of silt and sand grains. Therefore, the minimum aggregate size, $X_{min}$ (Eq.(16)) is reduced to the mean size of silt and sand grains ($x_n$), and the maximum aggregate size ($X_m$) is obviously reduced to the maximum sand grain size.

**Equation for the *K* Ratio and General Algorithm of Its Numerical Solution**

Using the above presentations of $(1-P_h)/(\rho_b U_h)$ (Eq.(3)), $V_i/V_a$ (Eq.(4)), $F(X,P_h)$ (Eq.(5)-(8)), and $X_{min}$ (Eq.(16)) one can present Eq.(1) and (2) as

$$K = \left\{ (1-\alpha)^3 - 3[\alpha/(1-\alpha)] \int_0^1 \frac{\eta^2 F(\eta, P_h) d\eta}{(\eta + [\alpha/(1-\alpha)])^4} \right\}^{-1} \tag{18}$$

where

$$\alpha \equiv x_n/X_m \;. \tag{19}$$

According to the physical meaning of $x_n$ and $X_m$ (see above) the $\alpha$ ratio can take values in the range $0 < \alpha < 1$. The case of $\alpha \to 0$ (in practice, $0 < \alpha \ll 1$) corresponds to high clay contents ($c \to 1$). Indeed [as noted above, see text after Eq.(10)] at high clay content $x_n$ strives to a clay particle size. That is, $x_n/X_m \ll 1$ and, therefore, $\alpha \ll 1$

(Eq.(19)). One can see that in this case (when $\alpha \to 0$), according to Eq.(18), $K \to 1$ as it should be at $c \to 1$ [2]. The opposite case of $\alpha \to 1$ (but less than 1) corresponds to low clay contents ($c \to 0$, in practice $c \ll c^*$). Indeed [as noted above, see text after Eq.(10)] at low clay content $x_n$ strives to a silt-sand grain size, and $X_m$ strives to the maximum sand grain size. That is, the maximum possible value of the $x_n/X_m$ ratio is unity (at a special possible grain-size distribution), and, therefore, $\alpha \to 1$ (Eq.(19)). One can see that in this case (when $\alpha \to 1$), according to Eq.(18), $K \to \infty$ (in practice $K \gg 1$) as it should be at $c \to 0$ and transition to a rigid soil [2].

Accounting for Eq.(17) for $X_m$ one can write $\alpha$ (Eq.(19)) as

$$\alpha = [1 + (X_{mz}/x_n - 1)(u_h/u_z)^{1/3}]^{-1} \quad . \tag{20}$$

The $u_h/u_z$ ratio at given parameters $Y_z$, $W_h$, $\rho_s$, $c$, $P_z$, $k$, and $U_{lpz}$ depends on $K$ [2]. This means that $\alpha = \alpha(K)$ from Eq.(20) is a known function of $K$. In addition, $P_h = P_h(K)$ (Eq.(9)). Thus, Eq.(18) is one relative to $K$ as an unknown value. Given the above seven input parameters: $Y_z$, $W_h$, $\rho_s$, $c$, $P_z$, $k$, and $U_{lpz}$ and two additional input parameters, $X_{mz}$ and $x_n$ (instead of $K$) one can solve Eq.(18) to find the $K$ ratio of the shrinking soil. Note, that additional parameters ($X_{mz}$ and $x_n$) are generalized characteristics of aggregate sizes ($X_{mz}$) and soil texture ($x_n$).

In a practical calculation of $K$ from Eq.(18) with given $Y_z$, $W_h$, $\rho_s$, $c$, $P_z$, $k$, $U_{lpz}$ and additional parameters, $X_{mz}$ and $x_n$, the simplest general algorithm is as follows. One can increase a test $K$ value with a step $\Delta K$ (*e.g.*, 0.001) starting from $K=1$. Using the test $K$ value together with given $Y_z$, $W_h$, $\rho_s$, $c$, $P_z$, $k$, $U_{lpz}$ one can find $u_s$, $u_z$, $u_h$, and $U_i$ [2], then $P_h$ (Eq.(9)), $\alpha$ (Eq.(20), using the additional parameters, $X_{mz}$ and $x_n$), and, finally, the right part of Eq.(18) that we designate (at the test $K$ value) as $K'(K)$. Such calculation of the $K'(K)$ values is continued with increasing $K$ by step $\Delta K$ until $K < K'(K)$. If $K_1 < K'(K_1)$ and $K_2 = K_1 + \Delta K > K'(K_2)$ the true $K$ value (*i.e.*, the $K$ root of Eq.(18)) is $K_1 + \Delta K/2$. Decreasing $\Delta K$ one can increase the accuracy of the $K$ calculation. Thus, the $K$ ratio of a shrinking soil can be calculated from other input parameters, $Y_z$, $W_h$, $\rho_s$, $c$, $P_z$, $k$, $U_{lpz}$ and generalized characteristics of aggregate sizes, $X_{mz}$ and soil texture, $x_n$.

## MATERIALS AND METHODS

### Two Possible Ways to Validate the $K$ Ratio Estimation

To validate the approach to $K$ ratio estimation as applied to a soil, one needs the data set, $Y_z$, $W_h$, $\rho_s$, $c$, $P_z$, $K$, $k$, and $U_{lpz}$ (including $K$) that permits one to predict the soil reference shrinkage curve [1, 2], and, in addition, data on the soil texture and structure to independently estimate the mean size of soil solids, $x_n$ and the maximum soil aggregate size in the oven-dried state, $X_{mz}$. Then, two ways of the validation are possible.

The first, direct way includes the following steps:

(i) estimating the soil parameters $u_s$, $u_z$, $u_h$, and $U_i$ that correspond to the given data set, $Y_z$, $W_h$, $\rho_s$, $c$, $P_z$, $K$, $k$, and $U_{lpz}$ according to [1, 2];

(ii) estimating $P_h$ (Eq.(9)) and $\alpha$ (using Eq.(20) with the given $x_n$ and $X_{mz}$ values);

(iii) numerical solving Eq.(18) relative to $K$ with $F(\eta, P_h)$ from Eq.(5)-(8) and at found $P_h$ and $\alpha$ (for the general algorithm see the end of the previous section); and

(iv) comparison between the found $K$ root and the initially given $K$ ratio (that enters the data set to predict the reference shrinkage curve of the soil).

The second, opposite, way includes the following steps:

(i) this step coincides with (i) of the first way;

(ii) estimating $P_h$ (Eq.(9));

(iii) numerical solving Eq.(18) relative to $\alpha$ with $F(\eta, P_h)$ from Eq.(5)-(8) and at the given $K$ ratio;

(iv) estimating the $\alpha' \equiv x_n/X_{mz}$ ratio from Eq.(20) with the $u_h/u_z$ ratio (see step (i)) and found $\alpha$ as

$$\alpha' \equiv x_n/X_{mz} = \{1 + [(1-\alpha)/\alpha](u_h/u_z)^{-1/3}\}^{-1} \quad ; \tag{21}$$

(v) comparison between the above found $\alpha' \equiv x_n/X_{mz}$ and the ratio $x_n/X_{mz}$ corresponding to the initially and independently given $x_n$ and $X_{mz}$ values.

### Data Used

Currently, examples of the data sets for the reference shrinkage curve prediction ($Y_z$, $W_h$, $\rho_s$, $c$, $P_z$, $K$, $k$, and $U_{lpz}$) are only available from [1, 2]. These data sets relate to the experimental shrinkage curves of six soils from [9], a soil from [10], and eight soils from [11]. However, data that are connected with the $x_n$ and $X_{mz}$ values of the soils only are in [11]. For this reason we only consider the eight soils from [11]. Table **1** reproduces the input data

($Y_z$, $W_h$, $\rho_s$, $c$, $P_z$, $K$, $k$, and $U_{lpz}$) of the eight soils for the reference shrinkage curve prediction from [1, 2]. In addition, Table **1** contains the predicted parameters, $u_s$, $u_z$, $u_h$, and $U_i$ for the soils from [1, 2].

Work [11] contains the data for the eight soils on weight fractions ($p_1$, $p_2$, $p_3$, $p_4$, $p_5$) of soil solids from five size ranges. Table **2** reproduces the $p_1$, $p_2$, $p_3$, $p_4$, $p_5$ values. These data indicate the size range of the largest sand grains as >0.2 mm. The maximum sand grain size, $x_m$ is unavailable for the soils. Thus, the data *per se* are not sufficient to find the accurate $x_n$ and $X_{mz}$ values and, correspondingly, the accurate $x_n/X_{mz}$ ratio and, then, α from Eq.(20). For this reason we could not immediately use both of the above possible ways to validate the $K$ ratio estimation. Nevertheless, the data from Table **2** contain the sufficient information to obtain approximate $x_n$ and $X_{mz}$ values. We could use these approximate estimates of $x_n$ and $X_{mz}$ in frames of the second way after its modification.

**Modified Second Way to Validate the *K* Ratio Estimation**

The modified second way differs from the initial one only with respect to step (v) (see the text after Eq.(21)). In the modified variant we compare the found α′≡$x_n/X_{mz}$ (from $Y_z$, $W_h$, $\rho_s$, $c$, $P_z$, $K$, $k$, and $U_{lpz}$) not with an independent estimate of the $x_n/X_{mz}$ ratio that flows out of the soil texture and structure, but with upper and lower estimates of the ratio. Using the texture data from Table **2** and some reasonable considerations we can find these upper and lower estimates of the $x_n/X_{mz}$ ratio for the eight soils (instead of an accurate estimate for each soil).

The upper estimate of the $x_n/X_{mz}$ ratio for each soil was obtained as follows. We exclude the clay fraction, <0.002 mm (Table **2**). That gives the upper estimate of the mean solids size, $x_n$ as

$$x_{n\ up}=\{0.011p_2+0.35p_3+0.125p_4+[0.2+(x_m-0.2)]p_5\}/(p_2+p_3+p_4+p_5) \qquad (22)$$

(0.011, 0.35, and 0.125 mm are the mean sizes of 0.002-0.02, 0.02-0.05, and 0.05-0.2 mm ranges). Accounting for the maximum sand grain size, $x_m$ being somewhere between 1 and 2 mm (e.g., [14]), we take three possible values, 1, 1.5, and 2 mm for $x_m$ in Eq.(22). Since $x_{n\ up}$ from Eq.(22) corresponds to a silt and sand mixture (without clay particles) the maximum aggregate size $X_{mz}=x_m$. Then, the $x_{n\ up}/x_m$ ratio gives the upper estimate of the real $x_n/X_{mz}$ ratio (that corresponds to the initial soil with clay) because the clay addition decreases the $x_n$ value compared to $x_{n\ up}$ and increases the $X_{mz}$ value compared to $x_m$. That is, the upper estimate, $(x_n/X_{mz})_{up}$, of the $x_n/X_{mz}$ ratio is

$$(x_n/X_{mz})_{up}=x_{n\ up}/x_m \qquad (23)$$

with $x_{n\ up}$ from Eq.(22) and possible $x_m$=1, 1.5, and 2 mm.

The lower estimate of the $x_n/X_{mz}$ ratio for each soil was obtained as follows. We exclude the sand fraction, >0.2 mm (Table **2**). That gives the lower estimate of the mean solids size, $x_n$ as

$$x_{n\ low}=(0.001p_1+0.11p_2+0.35p_3+0.125p_4)/(p_1+p_2+p_3+p_4) \ . \qquad (24)$$

As an upper estimate of $X_{mz}$ we can take the same value as above, $X_{mz}=x_m$ (=1, 1.5, and 2 mm). Then, the $x_{n\ low}/x_m$ ratio gives the lower estimate of the real $x_n/X_{mz}$ ratio (that corresponds to the initial soil with the sand fraction >0.2 mm) since the addition of the sand fraction increases the $x_n$ value compared to $x_{n\ low}$ and can decrease $X_{mz}$ compared to the $x_m$ value. That is, the lower estimate, $(x_n/X_{mz})_{low}$, of the $x_n/X_{mz}$ ratio is

$$(x_n/X_{mz})_{low}=x_{n\ low}/x_m \qquad (25)$$

with $x_{n\ low}$ from Eq.(24) and possible $x_m$=1, 1.5, and 2 mm.

If the found α′≡$x_n/X_{mz}$ (Eq.(21)) gets into the interval between the independent lower and upper estimates ($(x_n/X_{mz})_{low}$ and $(x_n/X_{mz})_{up}$) this means that the approach to the $K$ ratio estimation is in compliance with available data and, in any case, does not contradict them.

**RESULTS AND DISCUSSION**

Table **3** shows the α≡$x_n/X_m$ and α′≡$x_n/X_{mz}$ values found for the eight soils according to the second way (points (iii) and (iv), respectively). Table **3** also shows the lower and upper estimates of the $x_n/X_{mz}$ ratio (Eq.(22)-(25)) that were found for the eight soils with $x_m$=1, 1.5, and 2 mm and using the $p_1$, $p_2$, $p_3$, $p_4$, $p_5$ fractions from Table **2**. In addition, Table **3** shows $P_h$ values relating to $W=W_h$, to illustrate the difference between those and similar values, $P_z$ relating to $W$=0 (Table **1**).

Comparing α′ (Table **3**) with the lower and upper estimates, $(x_n/X_{mz})_{low}$ and $(x_n/X_{mz})_{up}$ (Table **3**), one can see that for all the eight soils and three possible $x_m$ values the following inequality is fulfilled

$(x_n/X_{mz})_{low} < \alpha' < (x_n/X_{mz})_{up}$ . (26)

Thus, all the independently found lower and upper estimates ($(x_n/X_{mz})_{low}$ and $(x_n/X_{mz})_{up}$; Table **3**) for the soils are in the agreement with the estimates of $\alpha' \equiv x_n/X_{mz}$ (Table **3**) that were found from $\alpha$ (Table **3**; Eq.(21)), and $\alpha$, in turn, was found from Eq.(18) with data sets, $Y_z$, $W_h$, $\rho_s$, $c$, $P_z$, $K$, $k$, and $U_{lpz}$ including the $K$ ratio from Table **1**.

These results validate the approach to the $K$ ratio estimation using the additional parameters, $X_{mz}$ and $x_n$ that flow out of the soil aggregate-size distribution in the oven-dried state and soil texture, respectively. Accordingly, the $K$ ratio in the data set of input parameters for prediction of the reference shrinkage curve [1, 2] can be replaced with the $X_{mz}$ and $x_n$ values.

It is worth emphasizing that all other input parameters, $Y_z$, $W_h$, $\rho_s$, $c$, $P_z$, $k$, and $U_{lpz}$ as well as two additional parameters, $X_{mz}$ and $x_n$ can be found independently of an experimental shrinkage curve. This fact is of principle importance. It means that the model from [1, 2] permits prediction of the soil reference shrinkage curve before its measurement and based only on a number of elementary physical properties of inter- and intraaggregate soil structures.

## CONCLUSION

Chertkov [1, 2] introduced the new physical parameter of an aggregated soil - the ratio of an aggregate solid mass to the solid mass of an intraaggregate matrix. Using an experimental shrinkage curve (with a negligible crack volume contribution [1, 2]) this parameter can be easily found as a ratio of two characteristic water content values (without fitting) or as a fitted parameter. This work presents some supplemental consideration and results to [1, 2] to show that the aggregate/intraaggregate mass ratio, as a fundamental property of aggregated soil with any clay content, can be found independently of an experimental shrinkage curve based on a number of measured soil parameters. For that we derive an equation for the aggregate/intraaggregate mass ratio and show that this ratio can be predicted as a root of the equation. This equation is validated using data on eight soils. The measured physical soil parameters for finding the aggregate/intraaggregate mass ratio and predicting the reference shrinkage curve, include: oven-dried specific volume ($Y_z$), maximum swelling water content ($W_h$), mean solid density ($\rho_s$), soil clay content ($c$), oven-dried structural porosity ($P_z$), the lacunar factor that characterizes the rate of the lacunar pore volume change with water content ($k$), oven-dried lacunar pore volume ($U_{lpz}$), the maximum size of aggregates in the oven-dried state ($X_{mz}$), and the mean size of the soil solids (clay particles, silt, and sand grains) by their weight fractions ($x_n$).

## NOTATION

$c$      soil clay content, dimensionless
$c^*$      critical clay content, dimensionless
$F(X,P_h)$      cumulative aggregate-size distribution at a given structural porosity, $P_h$, dimensionless
$I_o(\eta)$      function from Eq.(7), dimensionless
$K$      ratio of aggregate solid mass to solid mass of intraaggregate matrix, dimensionless
$k$      lacunar factor, dimensionless
$P_h$      structural porosity of the soil at $W=W_h$, dimensionless
$P_z$      oven-dried structural porosity, dimensionless
$p_1,\ldots, p_5$      weight fractions of soil solids from five size ranges (see Table **2**), dimensionless
$U_h$      maximum specific volume of the intraaggregate matrix, dm$^3$ kg$^{-1}$
$U_i$      contribution of interface aggregate layer to the specific volume of soil aggregates, dm$^3$ kg$^{-1}$
$U_{lpz}$      lacunar pore volume in the oven-dried state, dm$^3$ kg$^{-1}$
$u(W)$      relative volume of intraaggregate matrix, dimensionless
$u_h$      relative volume of intraaggregate matrix at $W=W_h$, dimensionless
$u_s$      relative volume of solids of intraaggregate matrix, dimensionless
$u_z$      relative oven-dried volume of intraaggregate matrix, dimensionless
$V_a(X)$      volume of a separate aggregate of size $X$, m$^3$
$V_i(X, x)$      volume of the interface layer of the mean thickness, $x$ relating to a separate aggregate of size $X$, m$^3$
$W_h$      maximum swelling water content, kg kg$^{-1}$
$W$      soil water content, kg kg$^{-1}$
$X$      current aggregate size, m
$X_m$      maximum size of aggregates, m
$X_{min}$      minimum size of aggregates, m
$X_{mz}$      $X_m$ in oven-dried state, m
$X_{minz}$      $X_{min}$ in oven-dried state, m
$x$      mean thickness of interface layer of aggregates, m

| | |
|---|---|
| $x_m$ | maximum sand grain size, m |
| $x_n$ | mean size of soil solids, m |
| $x_{n\,low}$ | lower estimate of the mean solids size, $x_n$, m |
| $x_{n\,up}$ | upper estimate of the mean solids size, $x_n$, m |
| $(x_n/X_{mz})_{up}$ | upper estimate of the $x_n/X_{mz}$ ratio, dimensionless |
| $(x_n/X_{mz})_{low}$ | lower estimate of the $x_n/X_{mz}$ ratio, dimensionless |
| $Y_z$ | oven-dried specific volume of soil, dm$^3$ kg$^{-1}$ |
| $\alpha$ | ratio from Eq.(19), dimensionless |
| $\alpha'$ | ratio from Eq.(21), dimensionless |
| $\eta$ | parameter from Eq.(6), dimensionless |
| $\rho_b$ | dry bulk density, g cm$^{-3}$ |
| $\rho_s$ | mean density of solids, g cm$^{-3}$ |

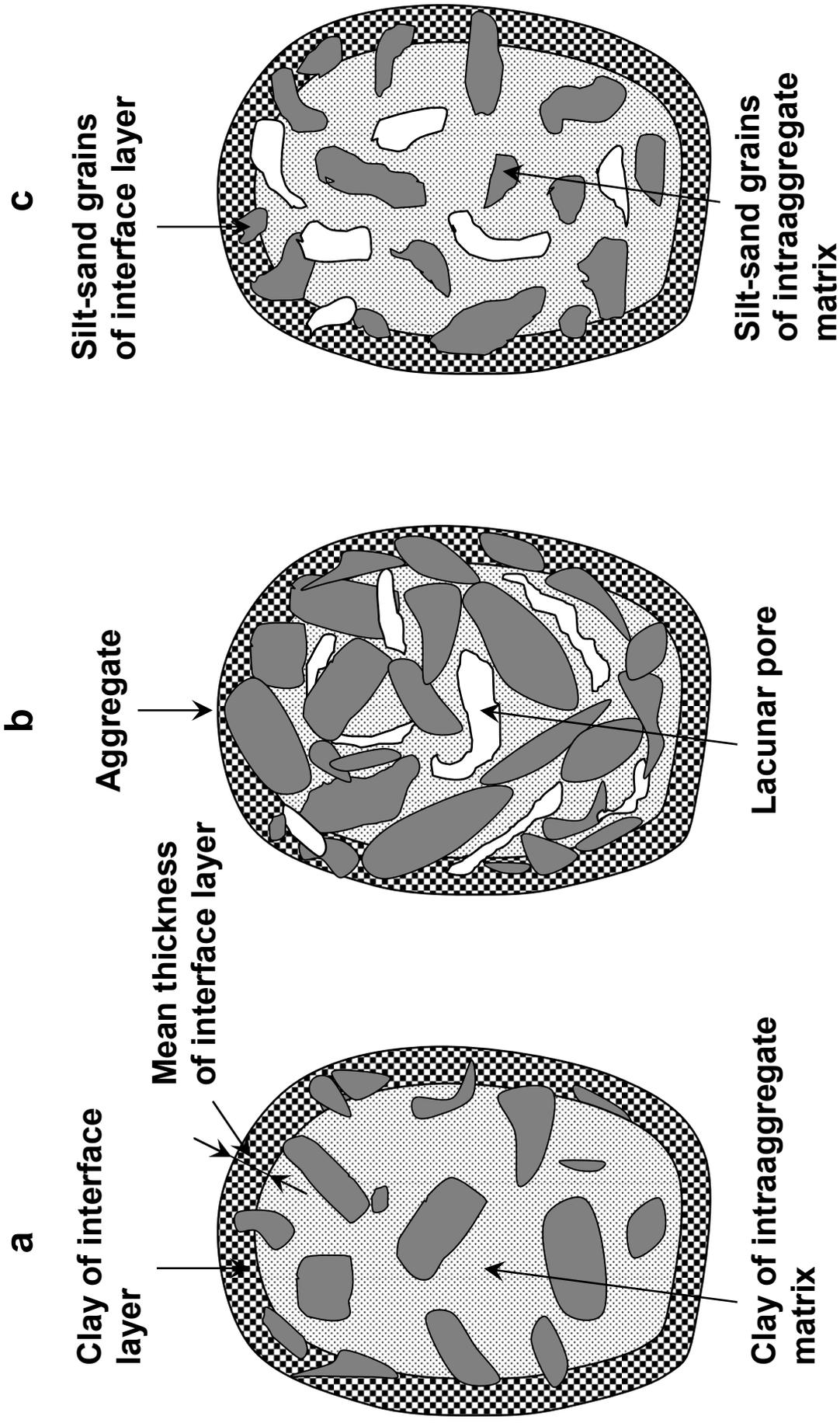

**Fig.1.** The illustrative scheme of the internal structure of aggregates at a clay content (reproduced from [2], Fig.**2**): (a) $c>c^*$, without lacunar pores; (b) $c<c^*$, with lacunar pores and silt-sand grain contacts at $W=0$; and (c) $c<c^*$, with lacunar pores and non-totally contacting silt-sand grains at any water content. $c^*$ is the critical soil clay content.

**Table 1. Input Parameters of the Eight Soils for Reference Shrinkage Curve Prediction and Predicted Parameters Participating in the $K$ ratio Estimation**[†].

| Data Source | | Soil Number | Input Parameters from [11] | | | | | Input Parameters of New Soil Characteristics from [1, 2] | | | | Predicted Parameters from [1, 2] | | | |
|---|---|---|---|---|---|---|---|---|---|---|---|---|---|---|---|
| | | | $Y_z$ | $W_h$ | $\rho_s$ | $c$ | $P_z$ | $K$ | $k$ | $U_{lpz}$ | $u_s$ | $u_z$ | $u_h$ | $U_i$ |
| | | | dm³ kg⁻¹ | kg kg⁻¹ | g cm⁻³ | | | | | dm³ kg⁻¹ | | | | dm³ kg⁻¹ |
| Ferruginous soil from Fig.3 of [11] | A horizon | 1 | 0.642 | 0.200 | 2.337 | 0.065 | 0 | 3 | 0.929 | 0.153 | 0.516 | 0.766 | 0.778 | 0.430 |
| | B1 horizon | 2 | 0.692 | 0.235 | 2.337 | 0.340 | 0.038 | 1.894 | 0.387 | 0.040 | 0.476 | 0.724 | 0.760 | 0.322 |
| | B2 horizon | 3 | 0.688 | 0.211 | 2.337 | 0.353 | 0.051 | 1.538 | 0.524 | 0.050 | 0.503 | 0.760 | 0.784 | 0.233 |
| | AB horizon | 4 | 0.682 | 0.229 | 2.337 | 0.187 | 0 | 2.499 | 0.771 | 0.130 | 0.483 | 0.751 | 0.782 | 0.416 |
| Ferralitic soil from Fig.3 of [11] | A horizon | 5 | 0.878 | 0.543 | 2.608 | 0.523 | 0 | 1.892 | 0.523 | 0.132 | 0.261 | 0.562 | 0.638 | 0.442 |
| | B1 horizon | 6 | 0.696 | 0.351 | 2.608 | 0.648 | 0 | 1.433 | 0.340 | 0.028 | 0.354 | 0.627 | 0.677 | 0.222 |
| | B2 horizon | 7 | 0.731 | 0.362 | 2.608 | 0.648 | 0.031 | 1.424 | 0.121 | 0.0074 | 0.346 | 0.625 | 0.673 | 0.222 |
| | AB horizon | 8 | 0.714 | 0.372 | 2.608 | 0.606 | 0 | 1.329 | 0 | 0 | 0.340 | 0.624 | 0.670 | 0.187 |

[†] $Y_z$, specific volume of an oven-dried soil; $W_h$, total gravimetric water content at maximum aggregate and soil swelling; $\rho_s$, mean density of solids; $c$, weight fraction of clay solids; $P_z$, oven-dried structural porosity of the soil; $K$, experimental estimate of aggregate/intraaggregate mass ratio; $k$, experimental estimate of lacunar factor value; $U_{lpz}$, experimental estimate of oven-dried value of the specific volume of lacunar pores; $u_s$, relative volume of all solids; $u_z$, relative volume of intraaggregate matrix at its shrinkage limit; $u_h$, relative volume of intraaggregate matrix at maximum swelling point; $U_i$, contribution of the interface aggregate layer to the specific volume of soil aggregates.

**Table 2. Data on the Texture of the Eight Soils from [11]†**

| Soil Number | $p_1$ | $p_2$ | $p_3$ | $p_4$ | $p_5$ |
|---|---|---|---|---|---|
| 1 | 0.065 | 0.063 | 0.245 | 0.490 | 0.138 |
| 2 | 0.340 | 0.097 | 0.203 | 0.288 | 0.089 |
| 3 | 0.353 | 0.112 | 0.189 | 0.239 | 0.120 |
| 4 | 0.187 | 0.088 | 0.228 | 0.392 | 0.112 |
| 5 | 0.523 | 0.163 | 0.089 | 0.142 | 0.083 |
| 6 | 0.648 | 0.080 | 0.108 | 0.116 | 0.047 |
| 7 | 0.648 | 0.080 | 0.108 | 0.116 | 0.047 |
| 8 | 0.606 | 0.128 | 0.071 | 0.146 | 0.049 |

† $p_1$, $p_2$, $p_3$, $p_4$, $p_5$, weight fractions of soil solids from size ranges, <0.002, 0.002-0.02, 0.02-0.05, 0.05-0.2, and >0.2 mm, respectively

**Table 3. Predicted α and α′ Ratios, Independent Lower and Upper Estimates of the α′ Ratio Found from Soil Texture[†], and the Structural Porosity, $P_h$ at the Maximum Swelling Point.**

| Soil Number | $\alpha \equiv x_n/X_m$ | $\alpha' \equiv x_n/X_{mz}$ | $x_m$ | $(x_n/X_{mz})_{low}$ | $(x_n/X_{mz})_{up}$ | $P_h$ |
|---|---|---|---|---|---|---|
| | | | mm | | | |
| 1 | 0.0994 | 0.0999 | 1 | 0.0818 | 0.1638 | 0 |
| | | | 1.5 | 0.0545 | 0.1338 | |
| | | | 2 | 0.0409 | 0.1188 | |
| 2 | 0.0725 | 0.0736 | 1 | 0.0480 | 0.1441 | 0.037 |
| | | | 1.5 | 0.0320 | 0.1180 | |
| | | | 2 | 0.0240 | 0.1049 | |
| 3 | 0.0485 | 0.0490 | 1 | 0.0426 | 0.1662 | 0.050 |
| | | | 1.5 | 0.0284 | 0.1411 | |
| | | | 2 | 0.0213 | 0.1286 | |
| 4 | 0.0808 | 0.0818 | 1 | 0.0650 | 0.1526 | 0 |
| | | | 1.5 | 0.0433 | 0.1245 | |
| | | | 2 | 0.0325 | 0.1105 | |
| 5 | 0.0538 | 0.0560 | 1 | 0.0253 | 0.1519 | 0 |
| | | | 1.5 | 0.0169 | 0.1303 | |
| | | | 2 | 0.0126 | 0.1195 | |
| 6 | 0.0288 | 0.0296 | 1 | 0.0208 | 0.1349 | 0 |
| | | | 1.5 | 0.0139 | 0.1123 | |
| | | | 2 | 0.0104 | 0.1009 | |
| 7 | 0.0375 | 0.0384 | 1 | 0.0208 | 0.1349 | 0.029 |
| | | | 1.5 | 0.0139 | 0.1123 | |
| | | | 2 | 0.0104 | 0.1009 | |
| 8 | 0.0224 | 0.0230 | 1.5 | 0.0159 | 0.1079 | 0 |
| | | | 2 | 0.0120 | 0.0965 | |

[†] α, predicted ratio of the mean soil solids size, $x_n$ to the maximum aggregate size, $X_m$ at the maximum swelling point; α′, predicted ratio of the mean soil solids size, $x_n$ to the maximum aggregate size, $X_{mz}$ in the oven-dried state; $x_m$, the maximum sand grain size; $(x_n/X_{mz})_{low}$, lower estimate of the $\alpha' \equiv x_n/X_{mz}$ ratio from soil texture; $(x_n/X_{mz})_{up}$, upper estimate of the $\alpha \equiv x_n/X_{mz}$ ratio from soil texture.